\documentclass[aps, prd, twocolumn, amsmath, floats,floatfix, superscriptaddress, nofootinbib,
showpacs]{revtex4}

\usepackage{amssymb}
\usepackage{amsmath}
\usepackage{verbatim}
\usepackage{mathrsfs}
\usepackage{amsfonts}
\usepackage{latexsym}
\usepackage{epsfig}
\usepackage{color}
\usepackage{graphicx,subfigure}
\usepackage{units}
\usepackage{envmath}
\usepackage{natbib}
\usepackage{ctable}
\usepackage{soul}
\usepackage{hyperref}
\begin{document}

%%%%%%%%%%%%%%%%%%
%%%   MACROS   %%%
%%%%%%%%%%%%%%%%%%

\definecolor{orange}{rgb}{0.9,0.45,0}

\newcommand{\re}{\mbox{Re}}
\newcommand{\im}{\mbox{Im}}

\newcommand{\fdg}[1]{\textcolor{green}{FDG: #1}}
\newcommand{\miq}[1]{\textcolor{red}{MMT: #1}}
\newcommand{\tf}[1]{\textcolor{red}{#1}}
\newcommand{\nsg}[1]{\textcolor{blue}{NSG: #1}}

\def\CovDev{D}
\def\Res{{\mathcal R}}
\def\Gammaflat{\hat \Gamma}
\def\metricflat{\hat \gamma}
\def\Dflat{\hat {\mathcal D}}
\def\part_n{\partial_\perp}

%=== Definition for abbreviations ===
\def\Lie{\mathcal{L}}
\def\A{\mathcal{X}}
\def\Aphi{\A_{\phi}}
\def\hAphi{\hat{\A}_{\phi}}
\def\E{\mathcal{E}}
\def\Ham{\mathcal{H}}
\def\M{\mathcal{M}}
\def\R{\mathcal{R}}
\def\p{\partial}

\def\hg{\hat{\gamma}}
\def\hA{\hat{A}}
\def\hD{\hat{D}}
\def\hE{\hat{E}}
\def\hR{\hat{R}}
\def\hcA{\hat{\mathcal{A}}}
\def\hDelt{\hat{\triangle}}

\def\na{\nabla}
\def\dif{{\rm{d}}}
\def\non{\nonumber}
\newcommand{\erf}{\textrm{erf}}
%====================================

\renewcommand{\t}{\times}

\long\def\symbolfootnote[#1]#2{\begingroup%
\def\thefootnote{\fnsymbol{footnote}}\footnote[#1]{#2}\endgroup}

%%%%%%%%%%%%%%%%%
%%%   TITLE   %%%
%%%%%%%%%%%%%%%%%

\title{Fermion-axion stars: static solutions and dynamical stability}
 
\author{Fabrizio Di Giovanni}
\affiliation{Departamento de
  Astronom\'{\i}a y Astrof\'{\i}sica, Universitat de Val\`encia,
  Dr. Moliner 50, 46100, Burjassot (Val\`encia), Spain}
  
\author{Davide Guerra}
\affiliation{Departamento de
  Astronom\'{\i}a y Astrof\'{\i}sica, Universitat de Val\`encia,
  Dr. Moliner 50, 46100, Burjassot (Val\`encia), Spain}

\author{Simone Albanesi}
\affiliation{Dipartimento di Fisica, Universit\`a di Torino, via P. Giuria 1, 
	10125 Torino, Italy}
\affiliation{INFN Sezione di Torino, Via P. Giuria 1, 10125 Torino, Italy}

\author{Miquel Miravet-Ten\'es}
\affiliation{Departamento de
  Astronom\'{\i}a y Astrof\'{\i}sica, Universitat de Val\`encia,
  Dr. Moliner 50, 46100, Burjassot (Val\`encia), Spain}

\author{Dimitra Tseneklidou}
\affiliation{Theoretical Astrophysics, IAAT, Eberhard-Karls University of T{\"u}bingen, 72076 T{\"u}bingen, Germany }

%%%%%%%%%%%%%%%%
%%%   DATE   %%%
%%%%%%%%%%%%%%%%

%\today

%%%%%%%%%%%%%%%%%%%%
%%%   ABSTRACT   %%%
%%%%%%%%%%%%%%%%%%%%

\begin{abstract} 

We construct spherically-symmetric static solutions of the Einstein-Klein-Gordon-Euler system involving a complex scalar field governed by a periodic potential which emerges in models of axion-like particles, and fermionic matter modeled by a perfect fluid with a polytropic equation of state. Such solutions describe gravitationally bound composites of fermions and axions which we dub as fermion-axion stars. Sequences of pure axion-stars in the existence domain may show the presence of multiple stable branches depending on the value of the decay constant parameter in the potential; this reflects in the appearance of multiple islands of stability in the 2-dimensional parameter space of fermion-axion configurations. We investigate the domain of existence for three different values of the decay constant, identifying one or more regions of linear stability making use of a method we already employed in previous works. We confirm the results from the linear analysis performing fully non-linear numerical relativity evolutions. In this context we perform several numerical simulations to identify regions in the parameter space where unstable models face different fates: the collapse to a Schwarzschild black hole, the migration to a stable model and finally the dispersion of the scalar field together with the dilution of the fermionic matter, which approaches a static fermion star model with very low mass. This latter scenario was never observed in previous models without the periodic potential.

\end{abstract}

%%%%%%%%%%%%%%%%
%%%   PACS   %%%
%%%%%%%%%%%%%%%%

\pacs{
95.30.Sf, % relativity and gravitation
04.70.Bw, 
04.40.Nr, 
04.25.dg
}

%%%%%%%%%%%%%%%%%%%%%%
%%%   MAKE TITLE   %%%
%%%%%%%%%%%%%%%%%%%%%%

\maketitle

\vspace{0.8cm}

%%%%%%%%%%%%%%%%%%%%%%
\section{Introduction}
%%%%%%%%%%%%%%%%%%%%%%

Unvealing the nature of dark matter (DM) is one of the fundamental challenges in modern cosmology. Its existence finds support from a wide set of observational results, such as the measurements of galaxy rotation curves, gravitational lensing and the cosmic microwave background~\cite{Hinshaw:2012aka,Reid:2012sw,Hu:2000ke,Caldwell:2009ix,Blake:2011en,Hlozek:2014lca,Chluba:2019nxa,Abazajian:2019eic}. Several candidates have been proposed as constituents of DM, including macroscopic objects like primordial black holes~\cite{Zeldovich:1967,Hawking:1971,Chapline:1975,Carr:1975} and a zoo of hypothetical particles which are considered to lack electromagnetic interactions with baryonic matter, thus being invisible through electromagnetic observations. Among the most compelling particle DM candidates there is the axion, a pseudo-scalar (boson) particle which was introduced in order to solve the strong CP problem by Peccei and Quinn~\cite{Peccei:1977}, but that could play a role in cosmology~\cite{ABBOTT1983133,Borsanyi_2016,PRESKILL1983127,Marsh_2016}. Ultralight axion-like fields naturally arise also from string theory compactifications~\cite{Conlon_2006,Svrcek_2006}, serving as another theoretical prediction of their existence. Motivated by these theoretical studies, various experiments have been proposed or are currently ongoing to search for this family of particles in a wide mass range~\cite{Hook_2018,Irastorza:2018dyq,Graham:2015}.

Bosonic particles can clump together to form localized and coherently oscillating equilibrium configurations which resemble Bose-Einstein condensates~\cite{Sin:1992bg,Chavanis:2011cz}. These compact objects are known in the literature as boson stars~\cite{jetzer:1992}, and they may have astrophysically relevant masses when the mass of the bosons is lower than $10^{-11}$~eV. Since the pioneer works of Kaup~\cite{Kaup:1968zz} and Ruffini and Bonazzola~\cite{Ruffini:1969qy}, their characterization has been broadened including different potentials such as the self-interaction~\cite{Colpi:1986ye}, the solitonic~\cite{Friedberg:1987}, the KKLS~\cite{Kleinhaus:2005,Kleinhaus:2008} and the axionic potentials~\cite{Guerra_2019}, including charge~\cite{Jetzer:1989av}, rotation~\cite{Yoshida:1997qf, Schunck:1996he}, oscillating solitonic stars~\cite{Seidel:1991zh}, multi-field boson stars~\cite{Alcubierre:2018ahf,Jaramillo:2020rsv,sanchis2021multifield}, or even vector field (Proca stars~\cite{brito2016proca}). The interested reader is addressed to the reviews found in references~\cite{Schunck:2003kk, liebling2017dynamical}. The dynamics of these configurations have been extensively studied with full non-linear numerical relativity simulations~\cite{Seidel:1990jh,Balakrishna:1997ej,sanchis2019nonlinear,DiGiovanni:2020ror,Siemonsen:2020hcg}; their stability properties have been investigated in~\cite{Jetzer1989,Gleiser:1988ih,Lee:1988av}, and a formation mechanism called gravitational cooling has been proposed in~\cite{Seidel:1993zk} by Seidel and Suen, and in~\cite{Guzman:2004wj, Guzman:2006yc} in the Newtonian limit, and has been extended to the vector field case in~\cite{di2018dynamical}. All these studies confirmed the dynamical robustness of scalar-field stellar systems.

In this work we consider the novel class of boson stars firstly studied in the relativistic regime by~\cite{Guerra_2019}. The scalar field is governed by a periodic potential inspired by that of the QCD axion, which depends on two independent parameters, the axion mass $\mu$ and the decay constant $f_{a}$. If such axion-like particles exist and they could form such compact objects, it is natural to assume that objects made out of a mixture of axions and fermions might also exist in the Universe, either considering the formation from a primordial gas comprised of axions and fermions, or by the dynamical capture of axionic or fermionic particles by an already formed neutron or axion star. Macroscopic composites of fermion and bosons are known in the literature as fermion-boson stars~\cite{HENRIQUES1990511,valdez2013dynamical,brito2015accretion,Brito:2016,DiGiovanni:2021vlu}. The presence of bosonic matter in fermion stars can modify their physical properties and potentially be observed as discussed in~\cite{DiGiovanni:2021ejn,DiGiovanni2022} for neutron stars (NSs) and in~\cite{Sanchis-Gual:2022} for white dwarfs. Gravitational-wave emission from orbital mergers of fermion-boson stars have also been analyzed in~\cite{Bezares:2019}. Here we will investigate the properties of macroscopic objects made of fermions and axions which we dub as fermion-axion stars.

We construct equilibrium configurations of fermion-axion stars and explore different possible values of the axion decay constant assessing their stability properties both in the linear regime and through fully non-linear numerical relativity simulations. Linear stability analysis can be carried out studying the radial perturbation of the equilibrium configurations and evaluating the modes in the linearized equations, as in~\cite{Chandrasekhar1964} for fermion stars and in~\cite{Jetzer:1989av,Gleiser:1988ih,Lee:1988av} for boson stars. The linear perturbation analysis has not yet been applied to fermion-boson stars, but the linear stability has been studied in previous works~\cite{valdez2013dynamical,valdez2020fermion,DiGiovanni:2020frc,DiGiovanni:2021ejn} using a variation of the method developed by Henriques et al.~\cite{HENRIQUES1990511,HENRIQUES1990737} which consists in evaluating the gravitational mass and the number of bosonic and fermionic particle as functions of the two free parameters searching for critical points for these three physical quantities in the 2-dimensional parameter space. In this work we employ this method for fermion-axion models, and confirm the results of the linear analysis through non-linear numerical evolutions, and we present a detailed study of the different fates of the unstable models, identifying the regions in the parameter space where models collapse to black holes (BHs) or migrate to a stable configuration or face the dispersion of the scalar field leaving a very low-mass fermion star remnant.

The paper is organized as follows. In section~\ref{sec:formalism} we present the basic equations employed to obtain the evolution equations and the matter source terms. Section~\ref{sec:ID} addresses the construction of the static configurations and in section~\ref{sec:stability} we briefly describe the linear analysis method and we present two sequences of equilibrium models and illustrate how the critical points are found. The numerical framework for the evolutions is described in section~\ref{sec:numerics}, and the results are presented in section~\ref{sec:evolutions}. Finally we report the conclusions and final remarks in section~\ref{sec:conclusions}. We employ units such that the relevant fundamental constants are equal to one $(G=c=M_{\odot}=1)$. For details on how to recover the physical units for radius and time we address the reader to our previous work~\cite{DiGiovanni:2020frc}.

%%%%%%%%%%%%%%%%%%%%%%
\section{Formalism}
\label{sec:formalism}
%%%%%%%%%%%%%%%%%%%%%%
%%%%%%%%%%%%%%%%%%%%%%

Models of mixed stars, where fermionic and bosonic matter coexist and interact only through gravity, can be characterised by a total stress-energy tensor which is the sum of two independent contributions, one from a perfect fluid and one from a complex scalar field, in the form
\begin{eqnarray}\label{Tmunu}
T_{\mu\nu}&=& T_{\mu\nu}^{\rm{fluid}} + T_{\mu\nu}^{\phi} ,\\
T_{\mu\nu}^{\rm{fluid}}&=& [\rho(1+\epsilon) + P] u_{\mu}u_{\nu} + P g_{\mu\nu}, \\
T_{\mu\nu}^{\phi}&=& - g_{\mu\nu}\partial_{\alpha}\bar{\phi}\partial^{\alpha}\phi - V(|\phi|) \nonumber \\
 &+& (\partial_{\mu}\bar{\phi}\partial_{\nu}\phi+\partial_{\mu}\phi\partial_{\nu}\bar{\phi}) . 
\end{eqnarray}

The perfect fluid is defined by its rest-mass density $\rho$, its pressure $P$, its internal energy $\epsilon$ and its four-velocity $u^{\mu}$. The complex scalar field is specified by its potential $V(|\phi|)$, which we choose in this work to be
\begin{equation}\label{potential}
V(|\phi|) = \frac{2 \mu^2 f_{a}^2}{B} \left( 1 - \sqrt{1-4B\sin^2{\Bigg(\frac{|\phi|}{2f_{a}}\Bigg)}}\right),
\end{equation}
where the constant $B=\frac{z}{z+1}\approx 0.22$ where $z=m_{u}/m_{d}\approx 0.48$ is the mass ratio between up and down quarks (see~\cite{DiCortona2016}). The two independent parameters $\mu$ and $f_{a}$ represent the particle mass and the decay constant respectively. The bar in the previous equations denotes complex conjugation. The system of equations governing the dynamics is given by the Einstein equations $G_{\mu\nu}=8\pi T_{\mu\nu}$, by the conservation laws of the fermionic stress-energy tensor and of the baryonic mass
\begin{eqnarray} 
\nabla_{\mu}T^{\mu\nu}_{\rm{fluid}} = 0, \label{conservation_laws1} \\
\nabla_{\mu}(\rho u^{\mu}) = 0, \label{conservation_laws2} 
\end{eqnarray}
and by the Klein-Gordon equation
\begin{equation}
\nabla_{\mu}\nabla^{\mu} \phi= U(\phi) \phi \label{Klein-Gordon}
\end{equation}
for the complex scalar field. In the previous equations the symbol $\nabla$ represents the covariant derivative with respect to the 4-metric $g_{\mu\nu}$, and $U(\phi) = \frac{\partial V(\phi)}{\partial|\phi|^2}$. The system is closed by a suitable choice of an equation of state (EoS) for the fluid, which relates the pressure with the rest-mass density and the internal energy. In this work we consider a polytropic EoS for the equilibrium configurations, and an ideal fluid EoS for the evolutions to take into account possible shock-heating effects, yielding
\begin{equation}\label{EOS}
P= K \rho^{\Gamma} = (\Gamma-1)\rho\epsilon\,.
\end{equation}
where we fix the parameters $ K = 100$ and $\Gamma = 2$. 

Our framework for the evolutions is based on a numerical code~\cite{Montero:2012yr} which employs a spherically-symmetric metric in isotropic coordinates 
\begin{equation}\label{isotropic_metric}
ds^2 = -\alpha^2 dt^2 + \psi^4 \gamma_{ij} (dx^{i} + \beta^{i}dt)(dx^{j} + \beta^{j}dt),
\end{equation} 
where $x^{i}=\{r,\theta,\varphi\}$ are the spherical coordinates, $\alpha$ and $\beta^{i}$ are the lapse function and the shift vector respectively, $\psi = e^{4\chi}$ is the conformal factor, and $\gamma_{ij}$ is the spatial metric, which takes the form
\begin{eqnarray}
\gamma_{ij} dx^idx^j = a(r)dr^2 + b(r)r^2 (d\theta^2 + \sin{\theta}^2 d\varphi^2) \,,
\end{eqnarray}
which depends on two generic functions $a(r)$ and $b(r)$.

We employ the Baumgarte-Shapiro-Shibata-Nakamura (BSSN) formulation of Einstein's equations~\cite{nakamura1987general,Shibata95, Baumgarte98},
in its covariant formulation introduced by Brown~\cite{Brown:2009,Alcubierre:2010is}. In this formalism, the evolved quantities are the spatial metric $\gamma_{ij}$, the conformal factor $\chi$, the trace of the extrinsic curvature $K$, its traceless part $A_a = A^r_r$, $A_{b}=A^{\theta}_{\theta}=A^{\varphi}_{\varphi}$, and the radial component of the so-called conformal connection functions $\Delta^r$ (see~\cite{Shibata95, Baumgarte98} for definitions). In our simulations we employ the ``non-advective $1+$log'' gauge condition for the lapse function $\alpha$ and a variation of the Gamma-driver condition for the shift vector $\beta^r$. The interested reader is addressed to~\cite{Montero:2012yr} for further details regarding the BSSN evolution equations, gauge conditions, and the formalism for the hydrodynamic equations of our numerical code. 

Even if we do not report here the entire system of equations, we remind the reader that they involve matter source terms emerging from suitable projections of the stress-energy tensor onto the spatial metric, namely
\begin{align}
\mathcal{E}&= n^{\mu}n^{\nu}T_{\mu \nu}, \\
j_i&=-\gamma_{i}^{\mu}n^{\nu}T_{\mu \nu}, \\
S_{ij}&= \gamma_{i}^{\mu} \gamma_{j}^{\nu} T_{\mu \nu},
\end{align}
where $\gamma^{\mu}_{\nu} = \delta^{\mu}_{\nu} + n^{\mu}n_{\nu}$ are the projection operators on the spatial hypersurfaces, $n_{\mu}$ is the unit normal vector, and $\delta^{\mu}_{\nu}$ is the Kronecker delta.

In the case of fermion-axion stars, we can evaluate the contribution to the matter source terms from the fluid and from the scalar field separetely. The perfect fluid matter source terms read
\begin{align}
\mathcal{E}^{\rm{fluid}}&= \left[\rho\,(1 + \epsilon) + P \right] W^2 - P, \\
j_r^{\rm{fluid}}&= e^{4\chi} a \left[\rho\,(1+\epsilon) + P\right] W^2 v^{r}, \\
S_{a}^{\rm{fluid}}&=  e^{4\chi} a \left [\rho\,(1+\epsilon) + P\right]W^2  v^{r} + P, \\
S_{b}^{\rm{fluid}}&= P,
\end{align}
where $S_a = S^r_r$ and $S_b=S^{\theta}_{\theta} = S^{\varphi}_{\varphi}$, $W = \alpha u^{t}$ is the Lorentz factor and $v^r$ is the radial component of the fluid 3-velocity. Following the work~\cite{Sanchis-Gual:2015bh} we introduce two auxiliary variables
\begin{eqnarray}
\Pi &=& \frac{1}{\alpha}(\partial_{t} - \beta^r\partial_{r})\phi, \label{Pi_scalar} \\
\Psi &=& \partial_r \phi. \label{Psi_scalar}
\end{eqnarray}
 In this formalism the bosonic contribution to the source terms takes the form
\begin{align}
\mathcal{E}^{\phi}&=  \left( \bar{\Pi}\,\Pi + \frac{\bar{\Psi}\Psi}{e^{4\chi}a}\right) + V(|\phi|) \label{rhomat_phi}\\
j_r^{\phi}&= -  (\bar{\Pi}\Psi + \bar{\Psi}\Pi), \label{j_phi}\\
S_{a}^{\phi}&=  \left(\bar{\Pi}\,\Pi + \frac{\bar{\Psi}\Psi}{e^{4\chi}a}\right) - V(|\phi|) \\
S_{b}^{\phi}&=  \left(\bar{\Pi}\,\Pi - \frac{\bar{\Psi}\Psi}{e^{4\chi}a}\right) - V(|\phi|),
\end{align}
 and the Klein-Gordon equation~\eqref{Klein-Gordon} is now recast as a first-order system of linear equations, which reads
\begin{align}
\partial_t \phi & = \beta^r\partial_r \phi + \alpha \Pi, \\
\partial_t \Pi & = \beta^r \partial_r \Pi + \frac{\alpha}{a e^{4\chi}} \biggl[\partial_r\Psi +  \Psi \biggl(\frac{2}{r} - \frac{\partial_r a}{2a} + \frac{\partial_r b}{b} \nonumber\\
         &+ 2 \partial_r{\chi}\biggr)\biggr] + \frac{\Psi}{a e^{4\chi}} + \alpha K \Pi - \alpha U(\phi), \\
\partial_t \Psi & = \beta^r\partial_r \Psi + \Psi \partial_r \beta^r + \partial_r(\alpha \Pi).
\end{align}

Finally we report here the elliptic sector of Einstein equations, which provides a set of constraint equations, namely the Hamiltonian constraint and the momentum constraint, which read as
\begin{align}
\mathcal{H} & = R - (A_{a}^2 + 2 A_{b}^2) + \frac{2}{3} K^2 - 16\pi \mathcal{E} = 0, \label{Hamiltonian-constraint} \\
\mathcal{M}_{r} & = \partial_{r}A_{a} - \frac{2}{3} \partial_{r}K + 6A_{a}\partial_{r}\chi + \nonumber \\
		& (A_{a}-A_{b}) (\frac{2}{r} +\frac{\partial_{r}b}{b}) - 8\pi j_{r} = 0, \label{Momentum-constraint}
\end{align} 
where $R$ is the Ricci scalar.  

%%%%%%%%%%%%%%%%%%%%%%%%%%%%%%%%%%%%%%%%%%%%%%%%%%%%
\section{Initial Data} 
\label{sec:ID}
%%%%%%%%%%%%%%%%%%%%%%%%%%%%%%%%%%%%%%%%%%%%%%%%%%%%
To perform numerical evolutions, a mandatory step is to construct initial data that solve the constraint equations~\eqref{Hamiltonian-constraint} and \eqref{Momentum-constraint} to obtain the physical solutions of Einstein equations. In this context, we employ a spherically-symmetric metric in Schwarzschild coordinates
\begin{equation} \label{Schwarzschild_metric}
ds^2 = -\alpha(r)^2 dt^2 + \tilde{a}(r)^2 dr^2 + r^2 ( d\theta^2 + \sin{\theta}^2 d\varphi^2),
\end{equation}
where $\tilde{a}(r)$ and $\alpha(r)$ are two geometrical functions; for simplicity we use the same symbol $r$ for the radial coordinate, even though this is now a different coordinate than the one appearing in~\eqref{isotropic_metric}. The bosonic field is assumed to have an harmonic time dependence $\phi(t, r) = \phi(r) e^{-i\omega t}$ where $\omega$ is its eigenfrequency. Assuming a static fluid, $u^{\mu}=(-1/\alpha,0,0,0)$, Einstein's equations can be recast as ordinary differential equations (ODEs) which read
\begin{align} \label{s1}
\frac{d\tilde{a}}{dr} & = \frac{\tilde{a}}{2}\left(\frac{1-\tilde{a}^{2}}{r} +8\pi r \biggl[ \frac{\omega^{2}\tilde{a}^{2}\phi^{2}}{\alpha^{2}}+ \tilde{a}^{2} V(|\phi|) \right.\nonumber \\
		& \left. +\Psi^{2}+\tilde{a}^{2}\rho(1+\epsilon)\biggl]\right.\biggl),
\end{align}
\begin{align} \label{s2}
\frac{d\alpha}{dr} & = \frac{\alpha}{2}\left(\frac{\tilde{a}^{2}-1}{r} +8\pi r \biggl[\frac{\omega^{2}\tilde{a}^{2}\phi^{2}}{\alpha^{2}} - \tilde{a}^{2} V(|\phi|)\right.\nonumber \\
		& \left.+\Psi^{2}+\tilde{a}^{2}P\biggl] \right.\biggl),
\end{align}
\begin{align} \label{s3}
\frac{d\phi}{dr} = \Psi ,
\end{align}
\begin{align} \label{s4}
\frac{d\Psi}{dr} & = -\biggl[1+\tilde{a}^{2}-8\pi r^{2}\tilde{a}^{2} \biggl(V(|\phi|) \nonumber \\
	&+\frac{1}{2}(\rho(1+\epsilon)-P)\biggl)\biggl]\frac{\Psi}{r}-\frac{\omega^{2}\tilde{a}^{2}\phi^{2}}{\alpha^{2}}-\tilde{a}^{2}U(|\phi|)\phi,
\end{align}
\begin{align} \label{s5}
\frac{dP}{dr} = -[\rho(1+\epsilon)+P]\frac{\alpha^{\prime}}{\alpha},
\end{align}
where the prime indicates the derivative with respect to $r$. This system is closed by the EoS as in equation~\ref{EOS}.

\begin{figure}[t]
\begin{minipage}{1\linewidth}
\includegraphics[width=0.95\textwidth]{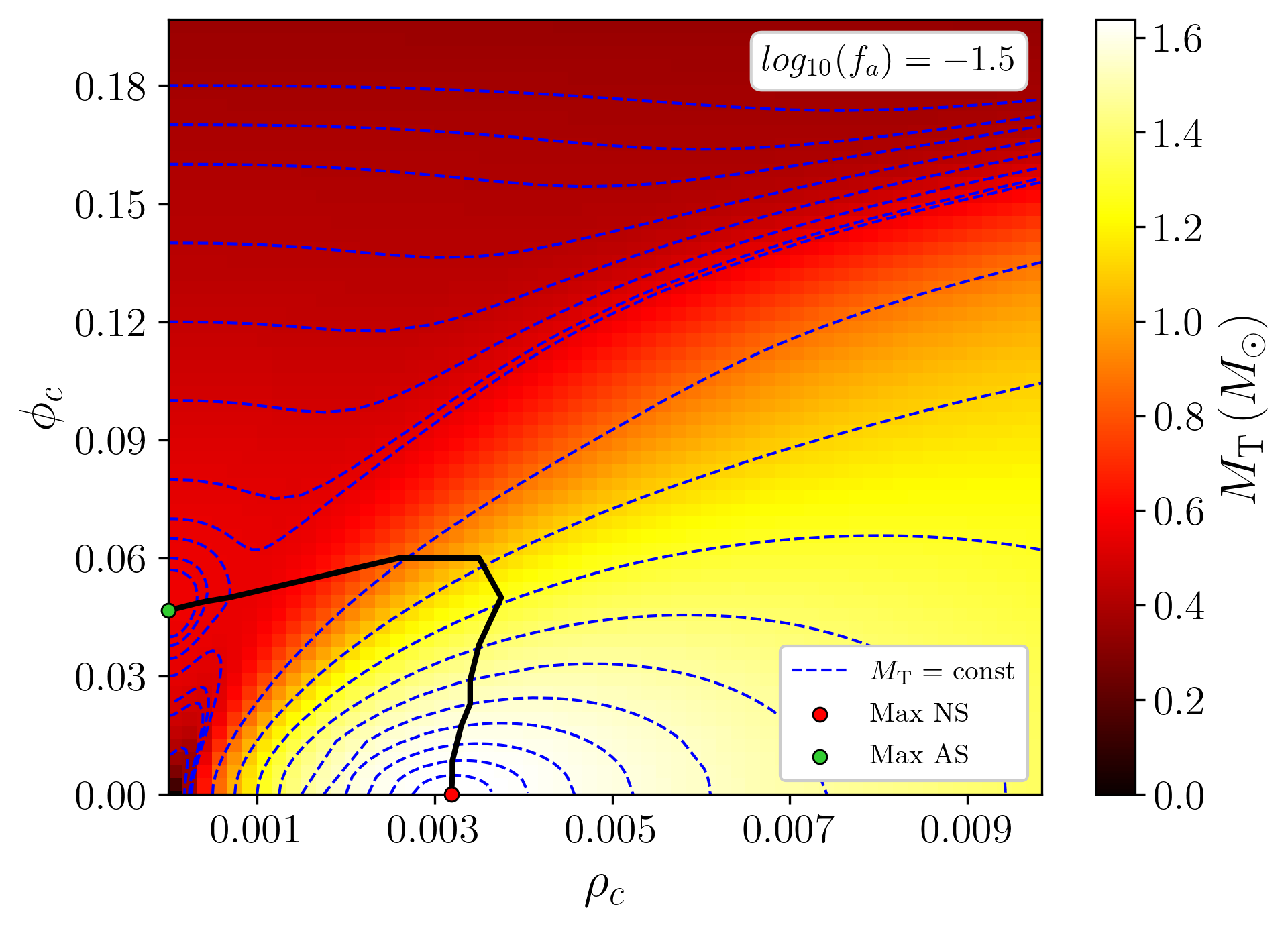}  
\includegraphics[width=0.95\textwidth]{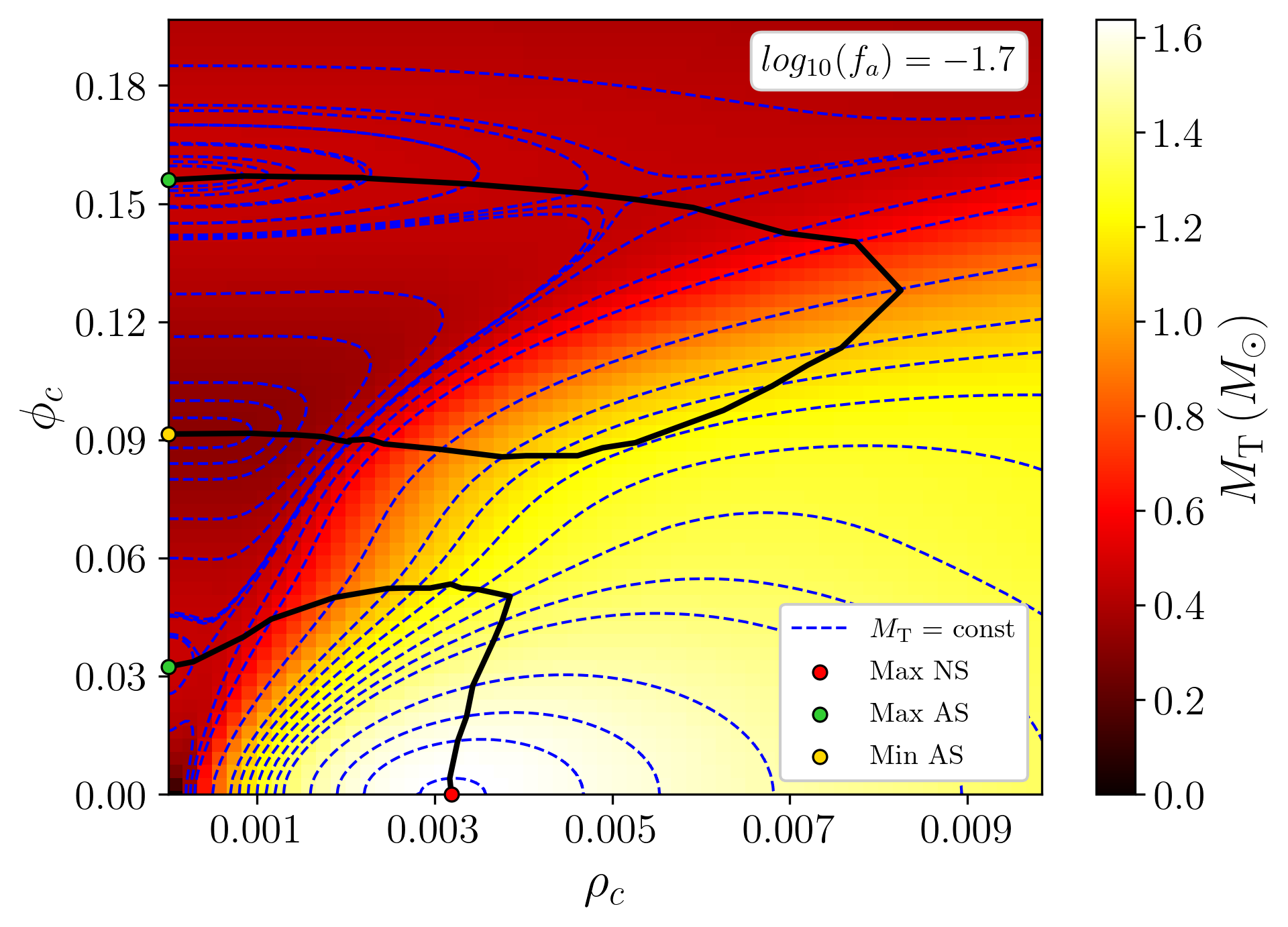} 
\includegraphics[width=0.95\textwidth]{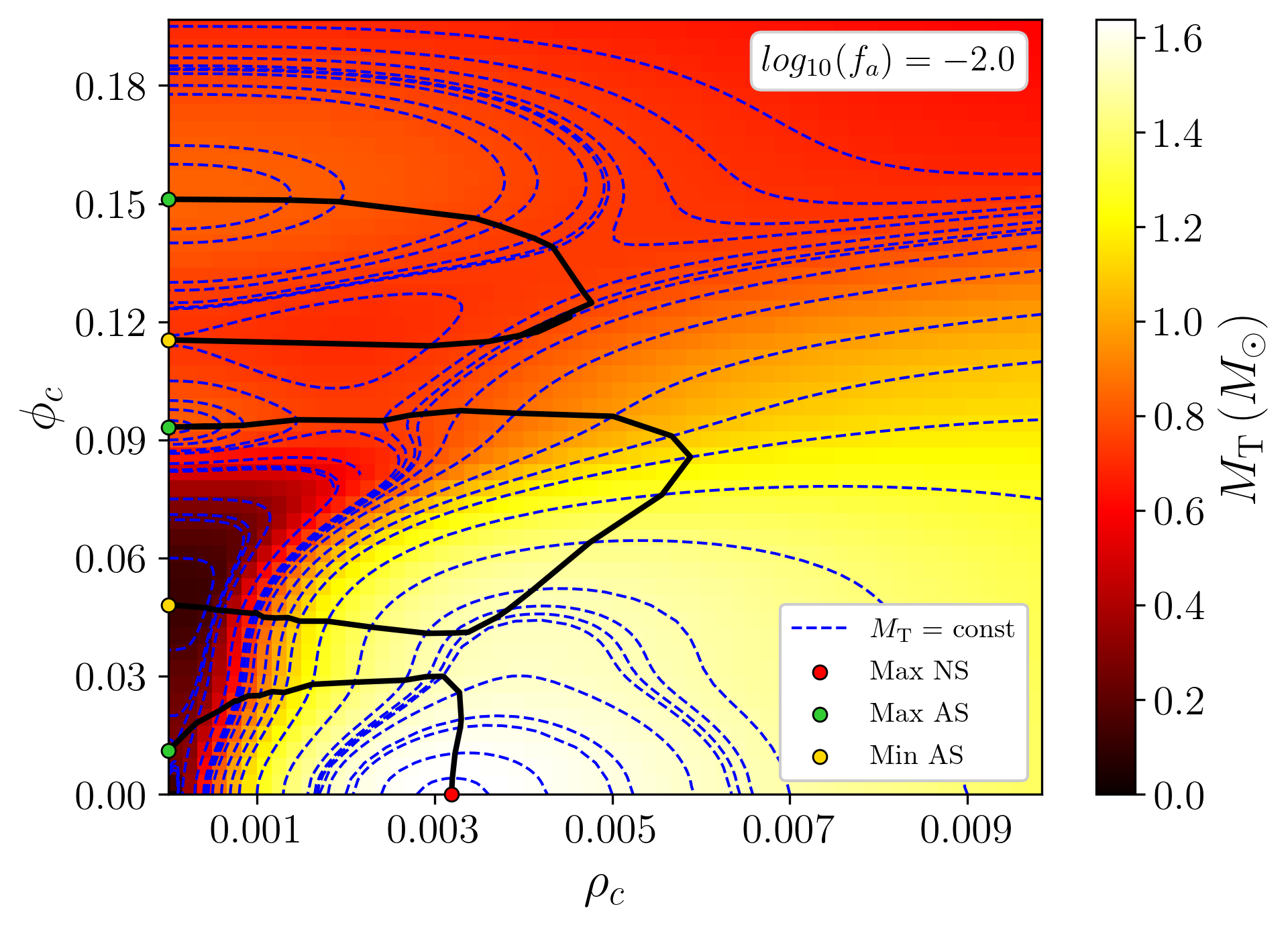} 
\caption{Equilibrium configurations of fermion-axion stars for $\log_{10}(f_{a})=-1.5$ (top), $\log_{10}(f_{a})=-1.7$ (middle), and $\log_{10}(f_{a})=-2.0$ (bottom). Dashed lines correspond to models with the same total mass $M_{T}$. The black solid line depicts the boundary between stable and unstable models.} 
\label{fig1} 
\end{minipage}
\end{figure}

To construct suitable equilibrium configurations we solve the ODE system with a 4th-order Runge-Kutta method, applying appropriate boundary conditions. Each numerical solution is characterised by the central values of the rest-mass density $\rho_{0}$ and of the scalar field $\phi_{0}$. We then require the metric functions to be regular at the origin, and we apply Schwarzschild outer boundary conditions. Finally we require that the scalar field vanishes at $r\rightarrow \infty$, and this condition can be fulfilled by evaluating the correct value of the eigenfrequency $\omega$; to achieve this, we make use of a two-parameter shooting method. To summarize, the set of boundary conditions that we apply are
\begin{eqnarray}
\label{eq:boundary_conditions}
&\tilde{a}(0) = 1, \hspace{0.3cm} & \phi(0) = \phi_{c}, \nonumber\\
&\alpha(0) = 1, \hspace{0.3cm} &  \lim_{r\rightarrow\infty}\alpha(r)=\lim_{r\rightarrow\infty}\frac{1}{\tilde{a}(r)},\nonumber\\
& \Psi(0)=0, \hspace{0.3cm} & \lim_{r\rightarrow\infty}\phi(r)=0, \nonumber\\
&\rho(0) = \rho_{c},  & \hspace{0.3cm}  P(0)=K\rho_{c}^{\Gamma}.
\end{eqnarray}

We evaluate the total gravitational mass for each model as
\begin{align} \label{mass}
M_T=\lim_{r\longrightarrow\infty}\frac{r}{2}\left(1-\frac{1}{\tilde{a}^2}\right),
\end{align}

which corresponds to the Anowitt-Desser-Misner (ADM) mass at infinity. 

%%%%%%%%%%%%%%%%%%%%%%%%%%%%%%%%%%%%%%%%%%%%%%%%%%%%
\section{Linear stability}
\label{sec:stability}
%%%%%%%%%%%%%%%%%%%%%%%%%%%%%%%%%%%%%%%%%%%%%%%%%%%%
\begin{figure}[t]
\begin{minipage}{1\linewidth}
\includegraphics[width=1.0\textwidth]{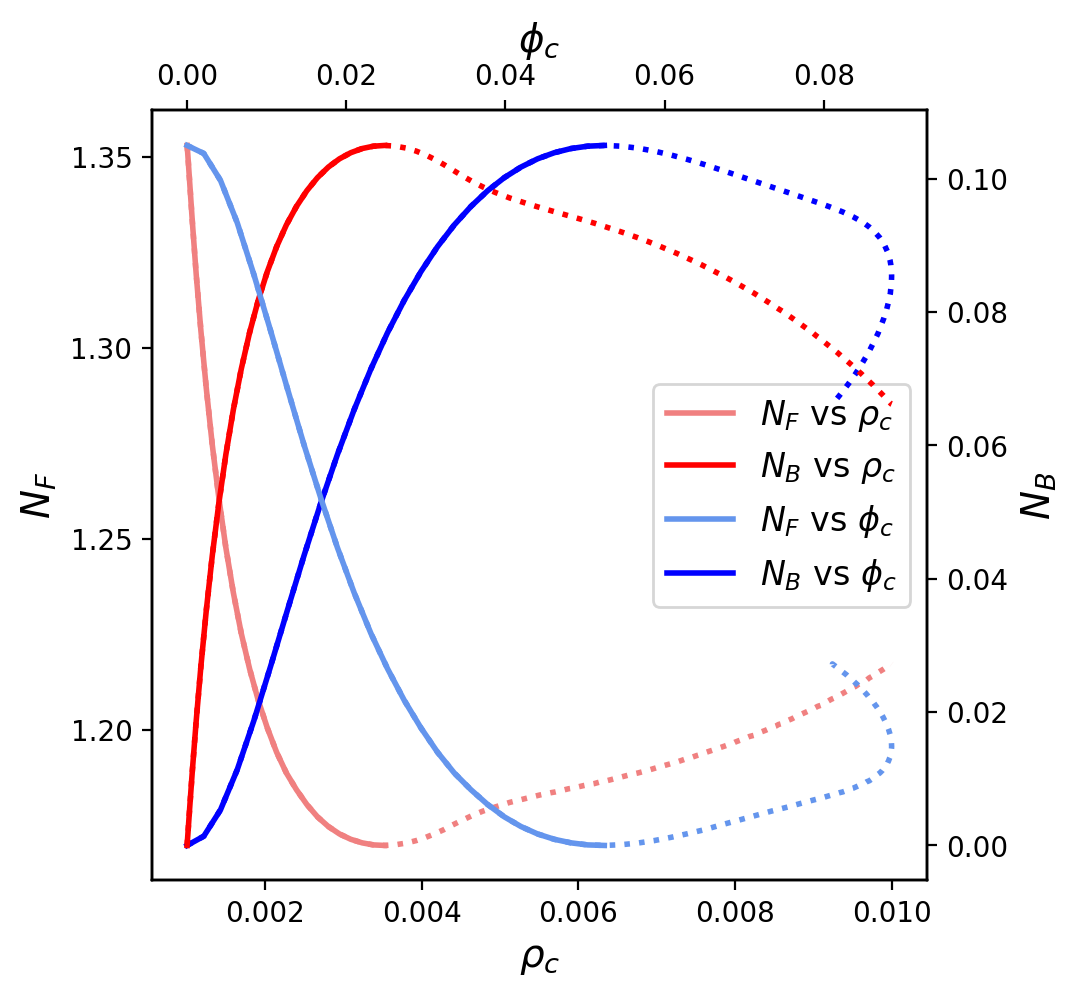}  \\
\includegraphics[width=1.0\textwidth]{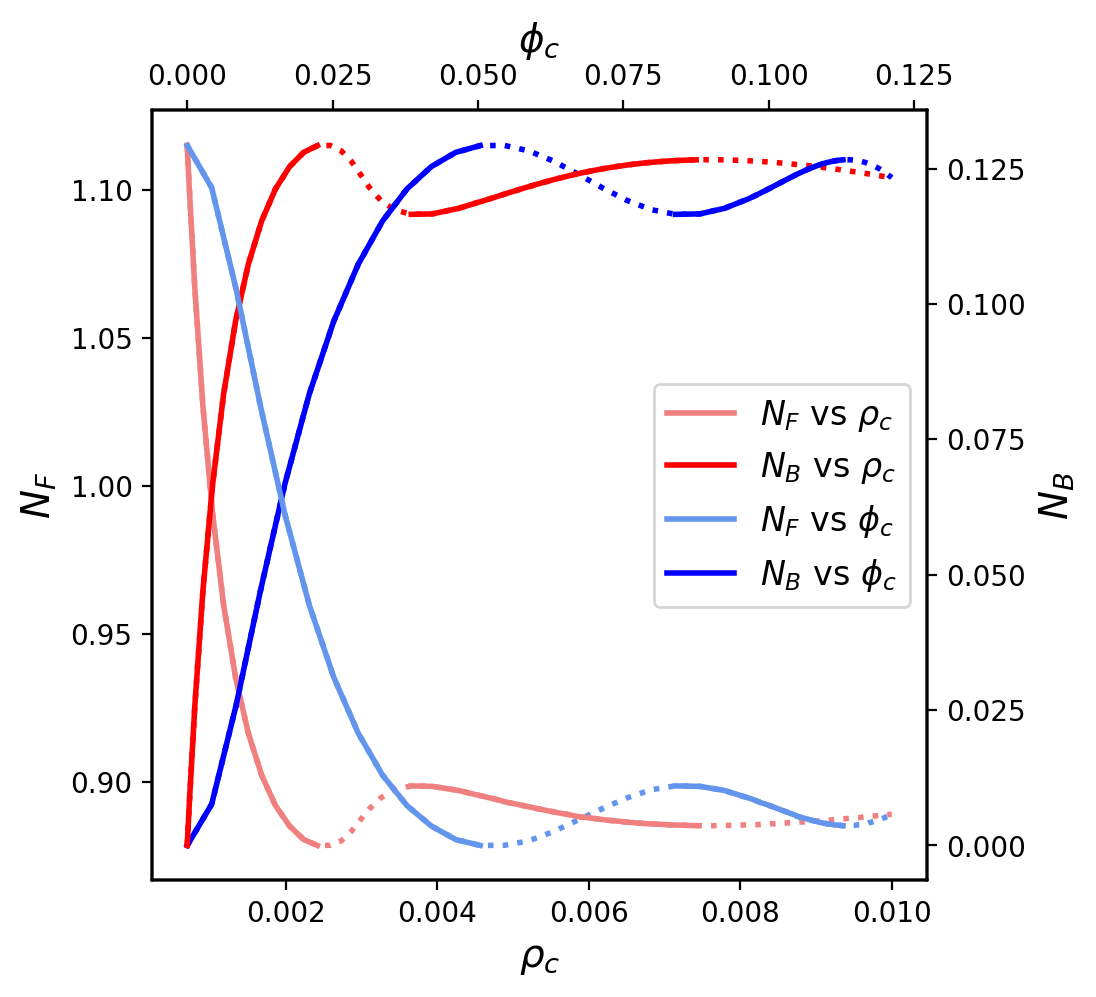}  

\caption{Number of fermions $N_{F}$ and bosons $N_{B}$ for the equilibrium configurations of equal mass $M_{T}=1.27$ (top plot) and $M_{T} = 1.06$ (bottom plot), varying the value of $\phi_{c}$ and $\rho_{c}$. The solid lines indicates the stable branches, while the dashed line the unstable ones.} 
\label{fig2} 
\end{minipage}
\end{figure}

In the previous section we have illustrated how to construct static solutions of fermion-axion stars. Once we have populated the  parameter space with models, a natural question that arises is whether we can delineate the boundary between the stable and unstable regions in such space. In this section we explain how to determine the linear stability of these solutions.

Identifying the stable and unstable branches for single-fluid systems like fermion stars and boson stars is straightforward, as this transition occurs at the equilibrium configuration with the largest mass, which is called the critical point. One method for computing the critical point is to consider an harmonic perturbation around each field static value and solve the linearized system of equations. This has been done for boson stars~\cite{Gleiser:1988ih,Lee:1988av,Jetzer1989,Kain:2021}, fermion stars~\cite{Chandrasekhar1964} and dark matter admixed NSs with fermionic dark matter~\cite{Comer1999,Kain2020}. We are not aware of such a study for fermion-boson or fermion-axion stars.

In the case of mixed systems, as we have a 2-dimensional space of parameters, the boundary between stable and unstable regions is not a point but it is a curve, called the critical curve. An alternative and simpler method to identify the critical curve for fermion-boson stars has been proposed by~\cite{HENRIQUES1990511}. Critical curves identify the transition from linearly stable and unstable with respect to perturbations which conserve mass and particle number, and hence fulfill the conditions
\begin{align}\label{criticalCurve}
\frac{\partial N_{B}}{\partial \rho_c} \Bigr|_{\substack{M=\rm{constant}}} = \frac{\partial N_{F}}{\partial \rho_c}\Bigr|_{\substack{M=\rm{constant}}} = 0, \nonumber \\
\frac{\partial N_{B}}{\partial \phi_c}\Bigr|_{\substack{M=\rm{constant}}} = \frac{\partial N_{F}}{\partial \phi_c}\Bigr|_{\substack{M=\rm{constant}}} = 0,
\end{align}
 where $N_{B}$ and $N_{F}$ are the number of bosonic particles and of fluid elements respectively. These quantities are associated with the conserved Noether charge related to the invariance under global $U(1)$ transformations $\phi\rightarrow\phi e^{i\delta}$ and with the conservation of the baryonic number respectively, and they can be evaluated by integrating their volume densities as follows
\begin{align}
N_B = 4 \pi \int \frac{\tilde{a} \omega \phi^2 r^2}{\alpha} dr, \hspace*{0.5cm} N_F= 4 \pi \int \tilde{a} \rho r^2 dr.
\end{align} 

To solve~\eqref{criticalCurve} for the critical curve, we follow the same procedure already presented in previous works~\cite{valdez2013dynamical,valdez2020fermion,DiGiovanni:2020frc}. We construct many contour lines with equal mass $M_{T}$ which populate the parameter space; we then move along each single line and we determine the point at which $N_{B}$ and $N_{F}$ present a critical value, meaning a change of the sign of their derivative with respect to the parameters $\rho_{c}$ and $\phi_{c}$. To automatize this procedure, we have developed a numerical code which can generate contour lines in the parameter space for fermion-boson stars, which details are briefly discussed in appendix~\ref{AppendixA}. This numerical code is publicly available\footnote{See the git repository at \href{https://github.com/SimoneA96/fermion-axion-pywrap}{https://github.com/SimoneA96/fermion-axion-pywrap}}.

In Fig~\ref{fig1} we depict with a colormap the total gravitational mass as a function of the two parameters characterizing the models $M_{T}(\rho_{0},\phi_{0})$ for 3 different values of the decay constant $\log_{10}(f_{a})=\{-1.5,-1.7,-2.0\}$, which show one, two, and three stable branches for axion stars respectively. On top of that, we show many contour lines of equal mass in dashed blue, and we construct with the method described in the previous paragraph the black solid line which is the boundary between the stable and unstable regions. Depending on the value of $f_{a}$ the existence line for axion stars present one or more critical points, outlining one or more stable branches~\cite{Guerra_2019}. We expected that the presence of more stable branches gives rise to different islands of stability for the fermion-axion configurations, and our results confirm this prediction. In the middle plot of figure~\ref{fig1} we observe a secondary region of linear stability that starts from the critical points of the axion star models, the minimum at $\phi_{c}=0.092$ and second maximum at $\phi_{c}=0.157$, and interestingly it extends up to around $\rho_{c}=0.008$ which is fairly higher than the value of the critical point for isolated NS which is around $\rho=0.0031$. Therefore we could reach stable neutron stars with extremely dense interiors due to the presence of axion particles, with interesting implications for the properties of dense matter. In the bottom plot we observe the appearance of a third island of stability corresponding to the third stable branch in the pure axion star existence plot.

We now focus on two illustrative examples of sequences of equilibrium models with $\log_{10}(f_{a})=-1.7$ with masses $M_{T} = 1.27$ and $M_{T}=1.06$ which start from a purely fermionic star; in figure~\ref{fig2} we depict the number of bosons $N_{B}$ and fermions $N_{F}$ of these two sequences as functions of $\phi_{c}$ and $\rho_{c}$. In the first case (top plot) we can identify only one critical point in the curve, corresponding to a maximum of $N_{B}$ and a minumum of $N_{F}$ at the values of $\rho_{c}=0.00352$ and $\phi_{c}=0.052$; this contour curve in fact only crosses the boundary of the primary stability region. All the models on the left of the critical point lie in the stable region (solid line),  and the ones on the right (dashed line) are unstable. The second case with $M_{T}=1.06$ (bottom plot) instead presents two stable branches, which correspond to the intervals in which $N_{B}$ ($N_{F}$) increases (decreases); this equal-mass curve crosses the primary stable region at the first maximum (minimum) of $N_{B}$ ($N_{F}$), then crosses the secondary stable region in the minimum (maximum) and second maximum (minimum) of $N_{B}$ ($N_{F}$). Sequences of equilibrium configurations which start from a purely fermion star have the feature that the number of fermions $N_{F}$ firstly decreases up to the critical point and the number of bosons $N_{B}$ increases; sequences starting with a pure axion star shows the opposite trend.

%TO BE FILLED WITH THE CORRECT VALUES
\begin{table*}[t!]
\caption{Static fermion-axion star models with decay constant $\log_{10}(f_{a})=-1.7$. From left to right the columns indicate the model name, its fate, the central value of the scalar field $\phi_c$ and of the rest-mass density $\rho_c$, the field frequency obtained with the shooting method $\omega_{\rm{shoot}}$, the normalized frequency $\omega$, the total Misner-Sharp mass $M_{T}$, the number of bosons to fermions ratio $N_{B}/N_{F}$, the number of bosons $N_{B}$, the radius containing $99\%$ of bosons, fermions and total particles, $R_{B}$, $R_{F}$, $R_{T}$, respectively. We used the Schwarzschild coordinates to evaluate the radii.}
\centering 
\begin{tabular}{c  c | c  c c c | c c c c c c}
\hline
\hline                  
Model & Fate  & $\phi_c$ & $\rho_{c}$  & $\omega_{\rm{shoot}}$ & $\omega$ & $M_{T}$ & $N_{B}/N_{F}$ & $N_{B}$ & $R_{B}$ & $R_{F}$ & $R_{T}$ \\ [0.5ex]
\hline

MS1 & Stable & $1.24\times 10^{-1}$ & $7.27\times 10^{-3}$ & 1.521 & 0.617 & 0.861 & 0.246 & 0.152 & 3.19 & 5.95 & 5.86 \\

MS2 & Collapsing & $1.11\times 10^{-1}$ & $7.78\times 10^{-3}$ & 1.485 & 0.885 & 1.060 & 0.143 & 0.127 & 3.26 & 6.24 & 6.18 \\

MS3 & Migrating & $1.45\times 10^{-1}$ & $8.62\times 10^{-3}$ & 1.741 & 0.550 & 0.595 & 0.670 & 0.181 & 3.00 & 4.51 & 4.36 \\

MS4 & Dispersing & $7.00\times 10^{-2}$ & $5.00\times 10^{-4}$ & 1.167 & 0.882 & 0.369 & 41.50 & 0.181 & 7.31 & 1.79 & 7.41 \\

\hline
\hline
\end{tabular}
\label{table1}
\end{table*}

%%%%%%%%%%%%%%%%%%%%%%%%%%%%%%%%%%%%%%%%%%%%%%%%%%%%
\section{Setup for evolutions} 
\label{sec:numerics}
%%%%%%%%%%%%%%%%%%%%%%%%%%%%%%%%%%%%%%%%%%%%%%%%%%%%

To confirm the study of the linear stability in the non-linear regime we perform numerical evolutions of the Einstein-Klein-Gordon-Euler system described in section~\ref{sec:formalism}, with the spherically symmetric numerical code developed by~\cite{Montero:2012yr}, upgraded with the evolution equations and the matter source terms of the complex scalar field by~\cite{Sanchis-Gual:2015lje}. We have extensively tested and used this numerical framework in past works (see e.g.~\cite{Sanchis-Gual:2015bh,Sanchis-Gual:2015sms,Sanchis-Gual:2016tcm,escorihuela2017quasistationary,Sanchis-Gual:2021phr}). The numerical code solves the Einstein equations in spherical isotropic coordinates, making use of a Partially Implicit Runge-Kutta (PIRK) method, developed in~\cite{Isabel:2012arx, Casas:2014}, to treat and handle the numerical instabilities coming from the terms in the equations that carry the typical $\frac{1}{r}$ singularities. We employ a non-equally spaced numerical grid firstly introduced in~\cite{Sanchis-Gual:2015sms}, which covers the computational domain with two different patches, a geometrical progression up to a certain radius and an hyperbolic cosine outside from it. This allows us to move the outer boundaries far away from the origin, and hence prevent the effects of reflections for longer computational time. 

In our simulations we consider a minimal resolution of $\Delta r=0.0125$, and a Courant-Friedrichs-Lewy factor of $\frac{\Delta t}{\Delta r}=0.3$. The grid is shifted by $\Delta r/2$ to avoid the origin, meaning that our inner boundary is set at $r_{min}=\frac{\Delta r}{2}$, while we set the outer boundary at $r_{max}=6000$. We adopt a $4$-th order Kreiss-Oliger numerical dissipation to our evolution equations to damp out high-frequencies modes. We employ an upwind scheme to treat the advection terms, and we impose radiative boundary conditions.

\begin{figure}[t]
\begin{minipage}{1\linewidth}
\includegraphics[width=1.0\textwidth]{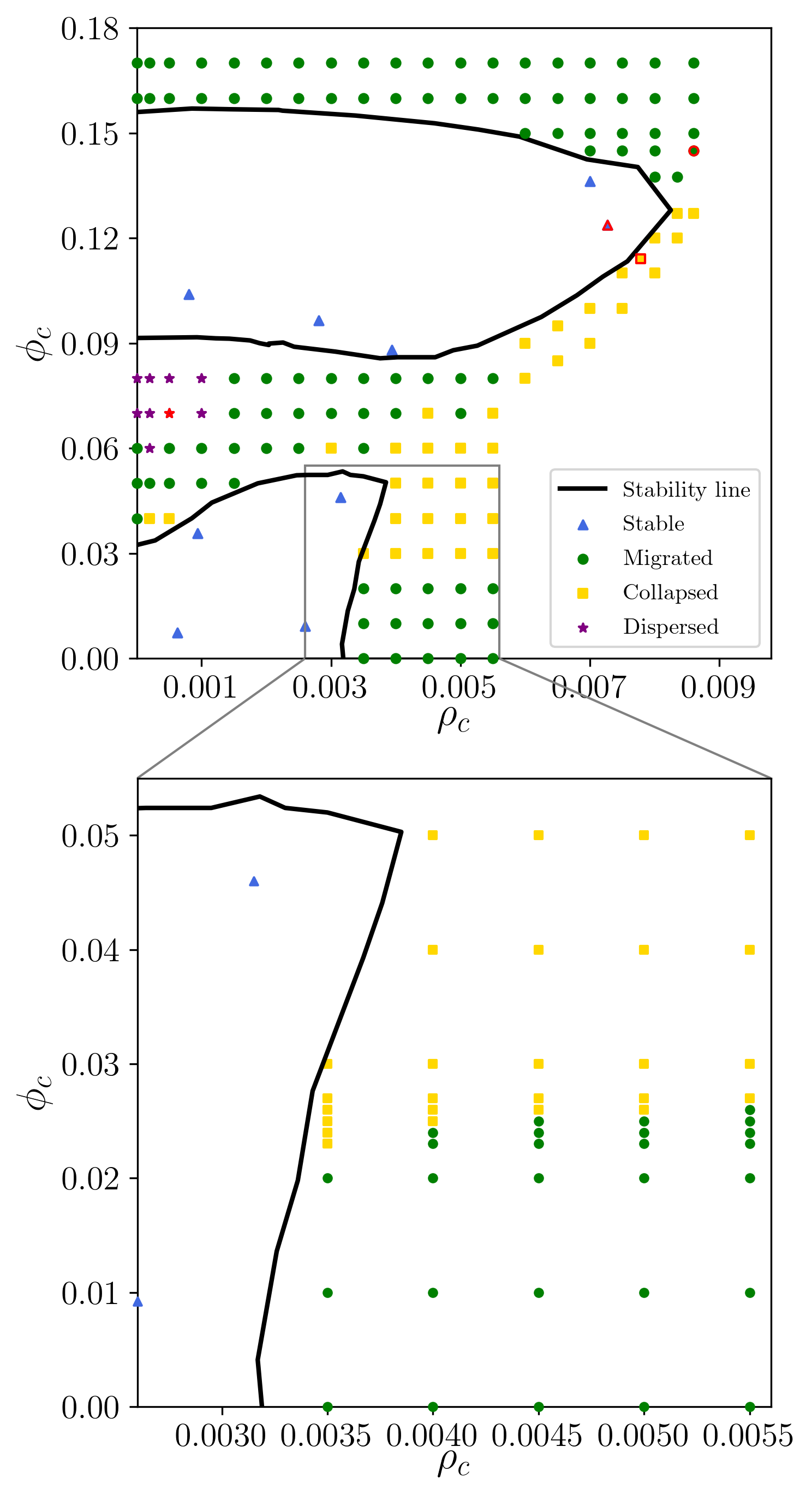}  
\caption{The thick black curve is the same critical curve as in middle panel of figure~\ref{fig1} for $\log_{10}(f_{a})=-1.7$. The blue triangles are linearly stable models that we evolved, the green dots are unstable models that migrates to the stable region, the yellow squares are unstable models which collapse to a BH, and the violet stars are models which show the dispersion of the scalar field. We highlight with a red outline the models whose physical properties are summarized in table~\ref{table1}. Bottom plot is a zoom of the region close to the unstable branch of pure NSs.} 
\label{fig3} 
\end{minipage}
\end{figure}

 %%%%%%%%%%%%%%%%
\section{Numerical evolutions}
\label{sec:evolutions}
%%%%%%%%%%%%%%%%

In this section we intend to verify if the regions of linear stability that we outlined in section~\ref{sec:ID} are populated by models which are also stable in the non-linear regime. To achieve this goal we perform numerical evolutions of the full non-linear Einstein-Euler-Klein-Gordon system, and we consider the equilibrium configurations to be weakly perturbed by the numerical truncation errors introduced by the discretization of the otherwise continuous differential equations. We expect that for stable mixed stars the fermionic density $\rho$ and the absolute value of the scalar field $|\phi|$ oscillate slightly around their initial values, while the scalar field itself oscillate at its eigenfrequency $\omega$.

For unstable models, however, even the small perturbation induced by the numerical discretization is expected to grow, and eventually the fate of these models can be of three types: the migration to a stable configuration, the gravitational collapse to a Schwarzschild BH or the total dispersion of the bosonic particles.

We perform numerical evolutions of several stable and unstable models for the three values of the decay constant $\log_{10}(f_{a})=\{-1.5 , -1.7 , -2.0 \}$, but we show the results only for $\log_{10}(f_{a})=-1.7$ as a representative example; the results from the other cases are similar.

\begin{figure*}[t!]
\centering

\includegraphics[width=1.0\textwidth]{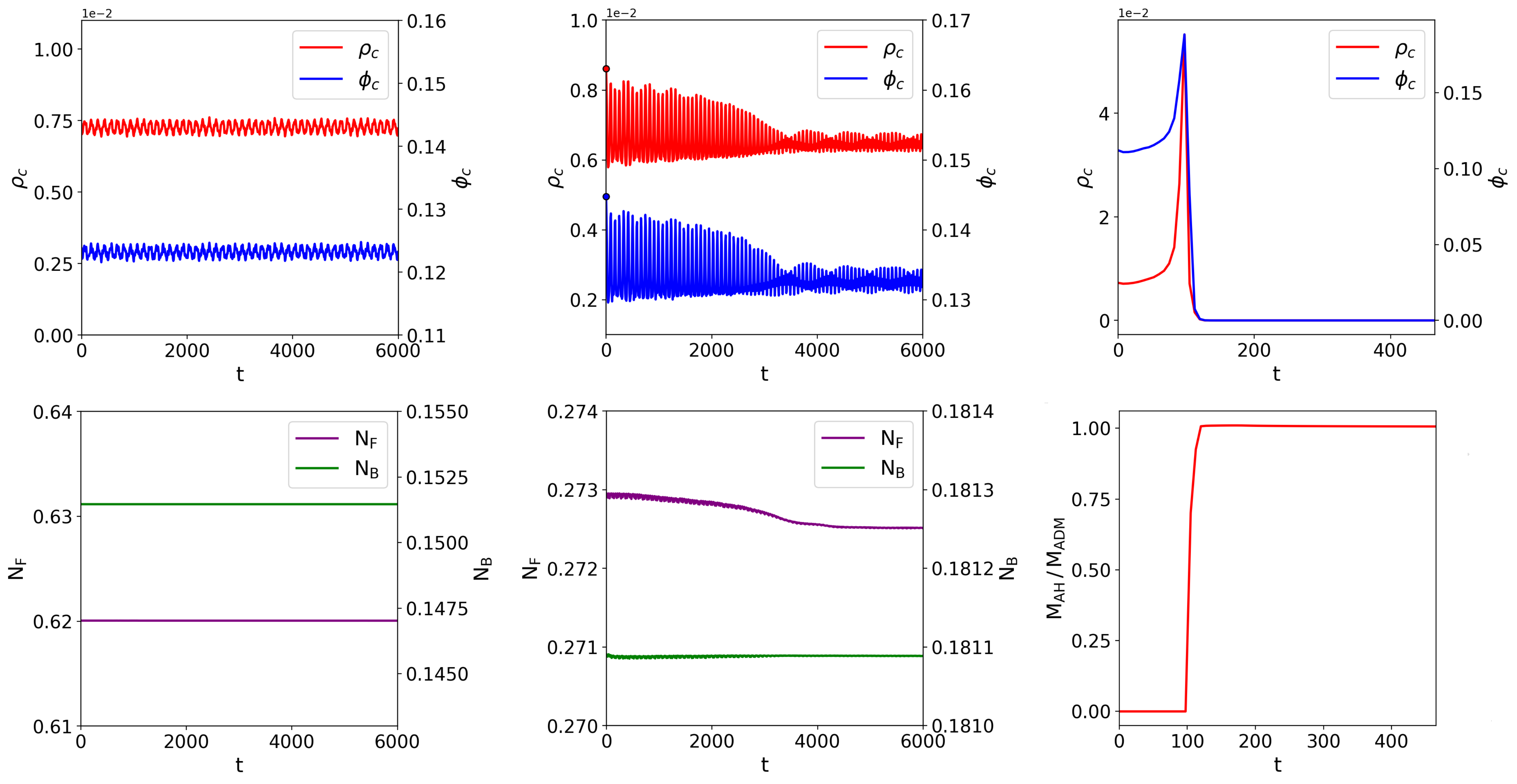} 

\caption{Time evolution of three different static models with decay constant $\log_{10}(f_{a})=-1.7$. Left panels show the central values of $\rho_{c}$ and $\phi_{c}$ (top row) and number of bosons $N_{B}$ and fermions $N_{F}$ (bottom row) for the stable model MS1. Middle panels illustrate the same quantities for the unstable model MS3 with the dots in the top panel corresponding to the initial values of $\rho_c$ and $\phi_c$. Right panels show $\rho_{c}$ and $\phi_{c}$ (top row) and the apparent horizon mass normalized with the ADM mass of the system (bottom row) for model MS2 which collapses to a Schwarzschild BH.}
\label{fig4}
\end{figure*} 

\begin{figure}[t!]
\centering
\includegraphics[width=0.4\textwidth]{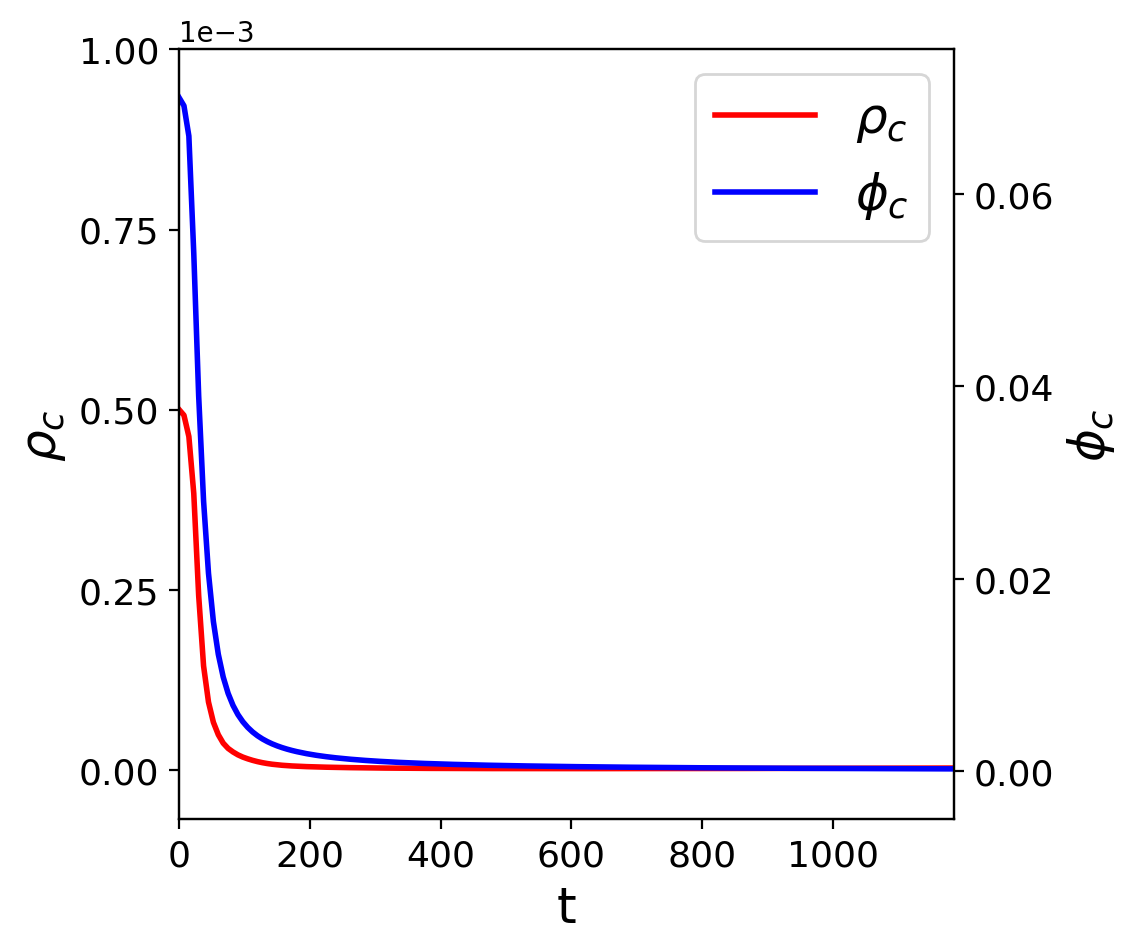} \\ \hspace{-0.95cm}
\includegraphics[width=0.33\textwidth]{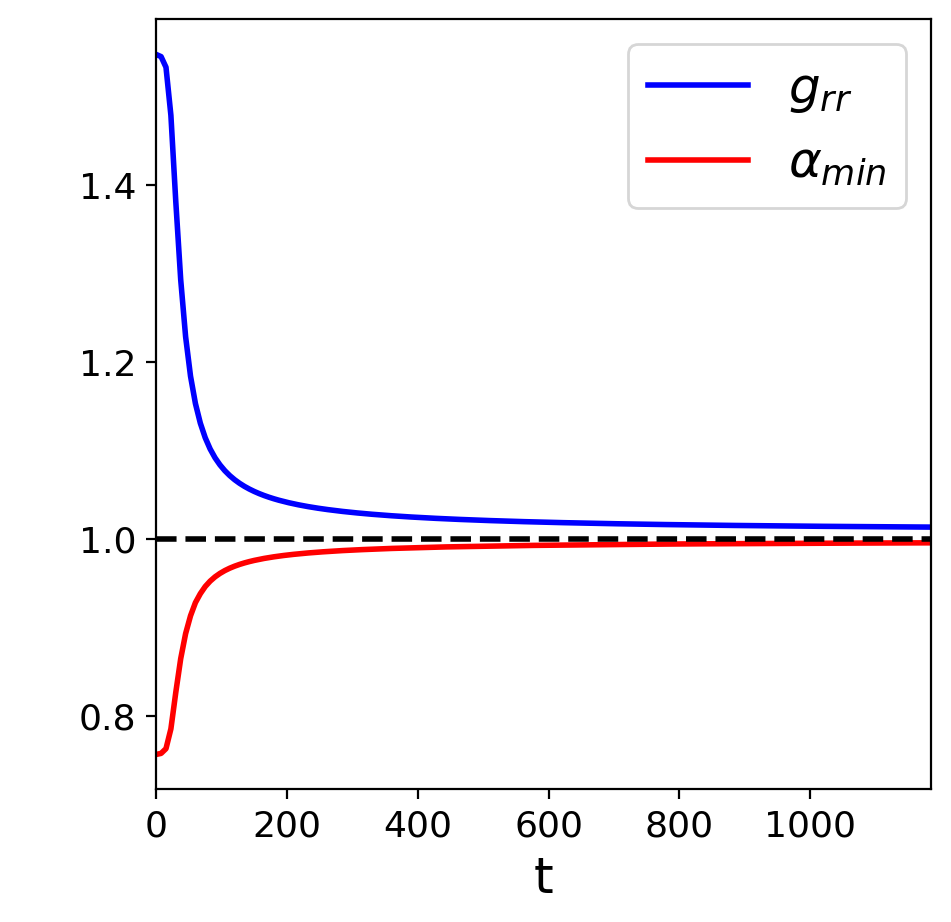} \\ \hspace{-0.85cm}
\includegraphics[width=0.34\textwidth]{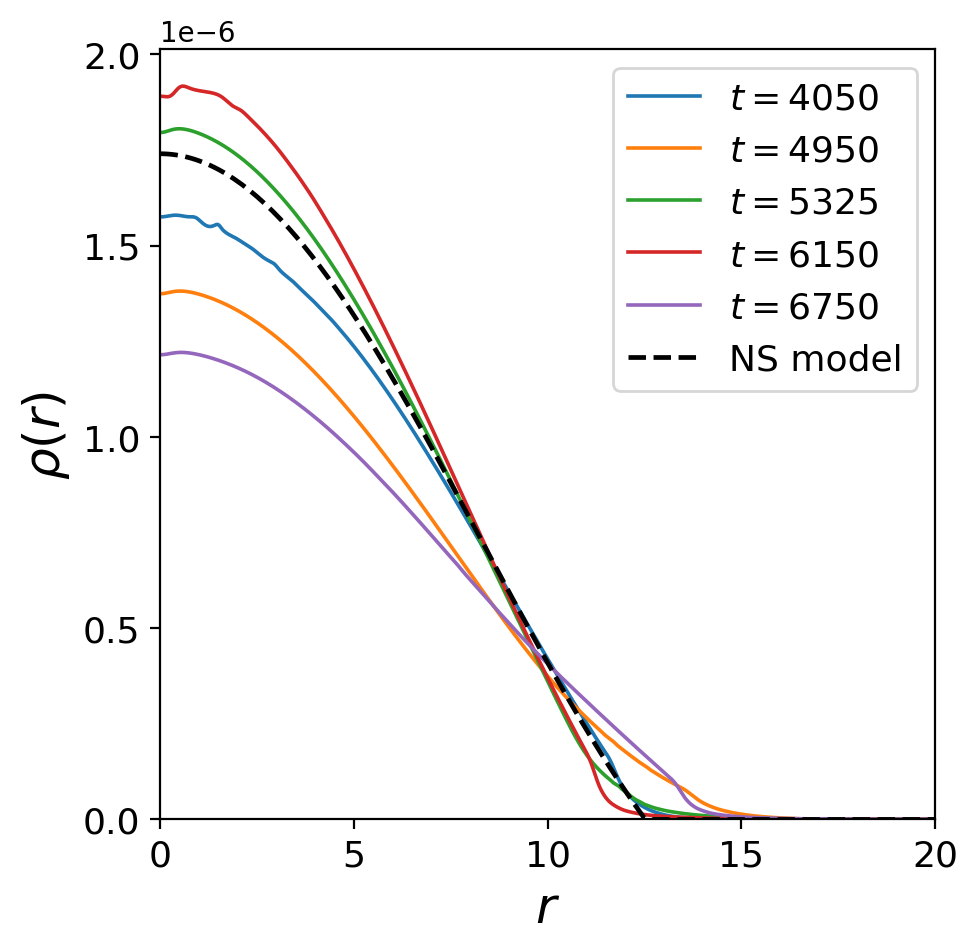} 
\caption{Time evolution of the dispersed model MS4 with decay constant $\log_{10}(f_{a})=-1.7$. Top plot illustrates the central value of the scalar field $\phi_{c}$ and the maximum value of the rest-mass density $\rho_{max}$ as a function of time. We notice that both these quantities approach zero indicating dispersion mechanism. In the middle plot we show the minimum value of the lapse $\alpha_{min}$ and the maximum value of the metric component $g_{rr}^{max}$ during the evolution. Both these quantities are converging to their Minkowski spacetime values, as expected. In the bottom plot we show the radial profile of latetime snapshots of $\rho$ compared with a static NS solution with a similar number of fermionic particles $N_{F}$.}
\label{fig5}
\end{figure}

Top plot of figure~\ref{fig3} shows the parameter space for $\log_{10}(f_{a})=-1.7$ populated by models that we have numerically evolved to verify the linear stability analysis. The blue triangles represent linearly stable models which we confirm to be stable in the non-linear regime, the green dots represent unstable models which migrate to the stable region, the yellow squares are unstable models which collapse to a Schwarzschild BH only being perturbed by the numerical truncation errors, and the violet stars represent models where we observe the dispersion of the scalar field. The area above the first stable region is mostly populated by models which migrates to the first stable island; close to the first unstable branch of pure axion stars we find models which show the dispersion of the scalar field. The region above the second stable island is only populated by models that migrate to the second stable region. The unstable branch of NS is populated by configurations that migrate; as we add a small amount of bosons to the system this feature is preserved, up to a certaint point where we discover a region where fermion-axion stars collapse to BHs. In the bottom plot of figure~\ref{fig3} we show a zoom of this area with more evolutions; we can appreciate how for higher values of $\rho_{c}$ we need a higher contribution of scalar matter to trigger the collapse to BH. In Table~\ref{table1} we report a list of the properties of one representative model for each of the possible fates, which we also depict in figures~\ref{fig4} and~\ref{fig5}.

We show in the left panels of figure~\ref{fig4} the time evolution of $\rho_{c}$ and $\phi_{c}$ (top plot) and the number of bosonic $N_{B}$ and fermionic particles $N_{F}$ (bottom plot) for the stable model MS1. We notice that the global quantities of the equilibrium solution are constant in time, revealing that the model is non-linearly stable. In the central panels of figure~\ref{fig4} we show the evolution of the same global quantities but for model MS3 which migrates to the stable region; the conserved quantities such as the number of fermionic and bosonic particles are constant during the evolution, but the system settles on a new static model, approaching the new central values $\rho_{c}=0.0064$ and $\phi_{c}=0.132$ which identify a point which lies on the secondary stable region. Finally in the right panels we show how in model MS2 the central values go to zero at the time when we also observe the appearance of an apparent horizon (bottom right plot), signaling the collapse to a BH. 

Model MS4 is illustrated in figure~\ref{fig5} where we depict in the top panel the time evolution of the central value of the scalar field $\phi_{c}$ and of the rest-mass density $\rho_{c}$ and in the middle panel the evolution of the minimum value of the lapse function $\alpha_{min}$ and the maximum value of the metric component $g_{rr}^{max}$; it can be appreciated that both the central value of the scalar field and of the rest-mass density drops to zero, while the metric components approach approximately the value 1 of the flat metric. This only happens for models which are very close to the first unstable branch of pure axion stars; interestingly while the scalar field is radiated away, the fermionic matter starts to get more dilute and there is a remnant object which seems to approach a static configuration of pure fermionic matter with a total ADM mass $M_{T}\approx 0.00435 \, M_{\odot}$. In the bottom plot of figure~\ref{fig5} we show a comparison between the latetime snapshots of the radial profile of $\rho$ and the static model; we can appreciate that the final configuration is oscillating approximately around this new configuration. Due to the low contribution of the fermionic component we can consider these models as pure axion stars which either accreted some baryonic matter, for example from a NS companion in a mixed binary system, or which formed from a primordial mixture of axionic and a small percentage of fermionic particles. A possible scenario to observe this phenomenon could be that of an axion star in the second stable branch which accreted a low amount of fermionic matter and which loses part of the axionic matter due to accretion onto a second more compact object, moving to the first unstable branch and triggering the dispersion mechanism. We point out that we present this result as an academic proof of concept, as we do not consider this scenario very likely to occur. Moreover, we describe the perfect fluid with a polytropic EoS with $\Gamma=2$ which is not a good description for such low rest-mass densities; a more precise study should involve more realistic EoS based on nuclear physics. 

%%%%%%%%%%%%%%%%%%%%%%%%%%%%%%%%%%%%%%%%%%%%%%%%%%%%%%%%%%%%%
\section{Conclusions}
\label{sec:conclusions}
%%%%%%%%%%%%%%%%%%%%%%%%%%%%%%%%%%%%%%%%%%%%%%%%%%%%%%%%%%%%%

We have studied models of fermion-axion stars, which are gravitationally bound objects composed by fermion and boson particles, where the latter are modeled by a complex scalar field whose equations of motion are governed by a periodic potential inspired by that of the QCD axion~\cite{Guerra_2019}. We have constructed equilibrium configurations for three different values of the decay constant $f_{a}$ and we have depicted the existence diagram in the parameter space spanned by the central rest-mass density and central scalar field amplitude. We have analyzed the linear stability and delineated the boundary between stable and unstable regions, being able to identify more than one island of stability as expected for those values of $f_{a}$ which show multiple stable branches in purely axion stars existence curves.

Finally we have presented a detailed study of the non-linear stability for a representative example. We have confirmed the results of the linear analysis; the evolutions of linearly stable models show how all physical quantities describing the star, such as the central values of the fields and the number of particles, remain constant in time. We have identified different areas in the unstable region where equilibrium models face different fates when they are weakly perturbed; some models migrate to the stable region, others collapse to a Schwarzschild BH, and finally we have found a small region close to the first unstable branch of pure axion stars in which the scalar field is rapidly dispersed away, and we find evidence of a remnant fermion star. This latter scenario was never observed in previous works on fermion-boson stars.
 
%%%%%%%%%%%%%%%%%%%%%%%%%%%%%%%%%%%%%%%%%%%%%%%%%

\acknowledgments
We thank Nicolas Sanchis-Gual and Jos\'e Antonio Font for useful comments and suggestions. This work was supported by the Spanish Agencia Estatal de Investigac\'ion (grant PGC2018-095984-B-I00), by the Generalitat Valenciana (PROMETEO/2019/071). FDG acknowledges support by the Generalitat Valenciana through the grant GRISOLIAP/2019/029. MMT acknowledges support by the Spanish Ministry of Universities (Ministerio de Universidades del Gobierno de Espa\~{n}a) through the FPU Ph.D. grant No. FPU19/01750. DG acknowledges support by the Spanish Ministry of Science and Innovation and the Nacional Agency of Research (Ministerio de Ciencia y Innovaci\'on, Agencia Estatal de Investigaci\'on) through the grant PRE2019-087617. We thank the Institute of Pure and Applied Mathematics (IPAM) at the University of California Los Angeles for hospitality during the Long Program "Mathematical and Computational Challenges in the Era of Gravitational Wave Astronomy", during which we started working on this project. 

%%%%%%%%%%%%%%%%%%%%%%%%%%%%%%%%%%%%%%%%%%%%%%%

\appendix
\section{Algorithm for equal-mass curves}\label{AppendixA}

In this appendix we discuss how we computed the equal-mass curves shown in blue 
in figure~\ref{fig1}. We recall that the computation of the mass $M_T$
associated to a pair $(\rho_c, \phi_c)$ involves the numerical solution of a set of ODEs.
In particular, we have to solve Eqs.~\eqref{s1}-\eqref{s5} using the boundary 
conditions of Eqs.~\eqref{eq:boundary_conditions}, and then to take the limit of Eq.~\eqref{mass}.
In order to accurately identify the stability region in the parameter space, we need to compute
many equal-mass curves, i.e. many level curves of $M_T(\rho_c, \phi_c)$.
To accomplish this task in an efficient and accurate way, we proceed as follows.
Given an initial point $\mathbf{p}_1=(\rho_c^1, \phi_c^1)$ in the parameter space, we compute
the corresponding mass $M_0 = M_T(\mathbf{p}_1)$. Then we find a second point
$\mathbf{p}_2=(\rho_c^2, \phi_c^2)$ of the level-curve identified by $M_0$ along 
a certain specified direction using a bisection algorithm requiring that 
$|M_0-M_T(\mathbf{p}_2)|<\epsilon$, where $\epsilon$ 
is a specified tolerance which we choose to be $\epsilon=10^{-6}$. The distance between $\mathbf{p}_1$ and the bisection interval used to find $\mathbf{p}_2$ is given
in input by the user.
Note that at this stage the algorithm 
can fail if the level-curve does not cross the bisection-interval. 
In our specific case, we always start from the $\rho_c$-axis or the $\phi_c$-axis, i.e. from the 
NS or axion-star case respectively, so that choosing the direction in which searching the 
second point is trivial. Having two points, we can apply the following procedure:
\begin{enumerate}
	\item we consider the line $\bar{r}_1$ passing through $\mathbf{p}_{n-1}$ and $\mathbf{p}_n$, then we consider a third point 
	$\mathbf{q}_*$ on $\bar{r}_1$ such that $d(\mathbf{p}_{n-1},\mathbf{p}_n)=d(\mathbf{p}_n,\mathbf{q}_*)$, where $d:{\mathbb R}^2\times {\mathbb R}^2 \rightarrow{\mathbb R}$ is the Euclidean distance;
	\item we consider $\bar{r}_2$, a line perpendicular to $\bar{r}_1$ that passes through $\mathbf{q}_*$, and we find 
	the two points $\mathbf{q}_L$ and $\mathbf{q}_R$ such that $d(\mathbf{q}_L,\mathbf{q}_*)=d(\mathbf{q}_*,\mathbf{q}_R)=d(\mathbf{p}_n,\mathbf{q}_*)$;
	\item we evaluate $M_T^L=M_T(\mathbf{q}_L)$ and $M_T^R=M_T(\mathbf{q}_R)$;
	\item depending on the sign of the product $(M_T^L-M_0)(M_T^R-M_0)$, we proceed as follows:
	\begin{enumerate}
		\item if $(M_T^L-M_0)(M_T^R-M_0)\leq 0$, then we apply the bisection algorithm on 
		the segment identified by $\mathbf{q}_L$ and $\mathbf{q}_R$ and we find the next point $\mathbf{p}_{n+1}$ requiring 
		$|M_0-M_T(\mathbf{p}_{n+1})|<\epsilon$. We call this method of finding $\mathbf{p}_{n+1}$
		the {\it tangent} method;
		\item If $(M_T^L-M_0)(M_T^R-M_0)>0$, i.e. if the level-curve is not passing through
		the segment identified by $\mathbf{q}_L$ and $\mathbf{q}_R$, then we build a square whose center is in $\mathbf{p}_{n}$, then
		we evaluate $M_T$ at the four vertices of the squares, and we search for the side crossed by
		the level-curve, and we apply the bisection algorithm
		to find $\mathbf{p}_{n+1}$ requiring 
		$|M_0-M_T(\mathbf{p}_{n+1})|<\epsilon$. We denote this method as the {\it square} method;
	\end{enumerate}
	\item we repeat this procedure until the curve closes or until we reach some specified boundary.
\end{enumerate}
Note that the square method guarantees to find a point, but it is slower than the tangent
method since requires two additional evaluations of $M_T$.

\begin{figure}[t!]
	\centering
	\includegraphics[width=0.22\textwidth]{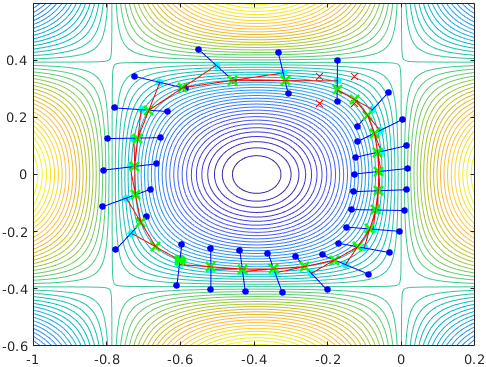}
	\includegraphics[width=0.22\textwidth]{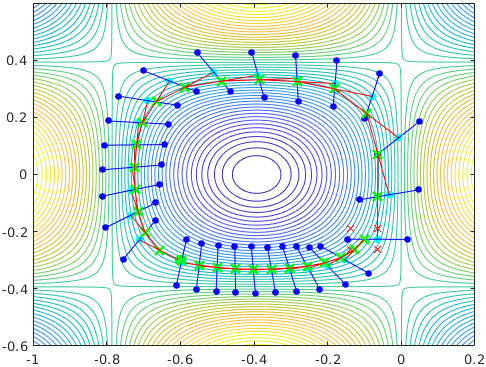}
	\includegraphics[width=0.22\textwidth]{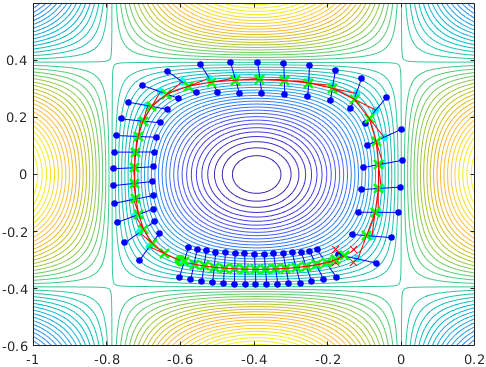}
	\includegraphics[width=0.22\textwidth]{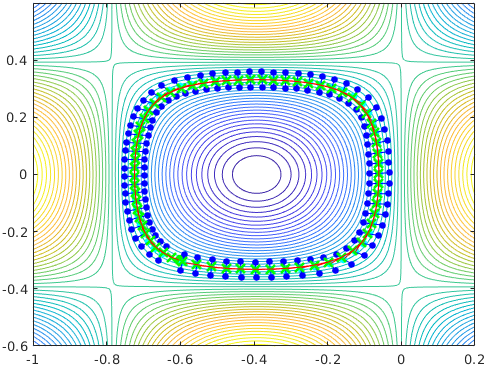}
	\caption{ 
		Level curves of the function $f(x,y)=\sin(4x)\cos(4y)$. On the background we show the ones compute with the \texttt{Matlab} function \texttt{countour()}, while on top we show
		the points of the level curves found by our algorithm (green points). We consider as initial point $\mathbf{p}_1=(-0.6, -0.30103)$,
		that corresponds to $f(\mathbf{p}_1)\simeq-0.242$, and we search the second point in the north direction using a tolerance of $\epsilon=10^{-10}$. We show the results for 4 different initial steps. The red lines are segments of $\bar{r}_1$, the blue lines are segment of $\bar{r}_2$, the
		cyan point is $\mathbf{q}_*$, the blue dots are  $\mathbf{q}_L$ and $\mathbf{q}_R$, and the red crosses are the vertex of the square created in the cases where the {tangent} method fails. 
	}
	\label{fig:analytic_lcurve}
\end{figure}

This method can be applied to any function $f(x,y)$, but if the function
is known in closed form then there are faster algorithms to find the corresponding level curves.
However, in order to test our algorithm, we consider an analytical function
and compare the contour plot produced by the \texttt{Matlab} function \texttt{countour()}
with the level-curve that we find with our algorithm.
An illustrative example is shown in figure~\ref{fig:analytic_lcurve}.
As can be seen, the distance between the points tends to increase up to a 
point where the tangent method fails and thus we have to find the next point using the square method. After this step, a relatively small distance between the points is restored. Note 
that this is not imposed in
the code, but it is just a consequence of the aforementioned procedure.
Finally, consider that using a small initial step almost always guarantees 
the success of the tangent method.

\bigskip

%\newpage

\bibliography{num-rel2}

\end{document}